\documentclass{article}

\usepackage[dblblindworkshop, final]{neurips_2026}
\workshoptitle{Economics for Machine Learning}

\usepackage[utf8]{inputenc}
\usepackage[T1]{fontenc}
\usepackage{url}
\usepackage{booktabs}
\usepackage{amsfonts}
\usepackage{nicefrac}
\usepackage{microtype}
\usepackage{xcolor}

\usepackage{amsmath}
\usepackage{amssymb}
\usepackage{graphicx}
\usepackage{amsthm}
\usepackage{soul}
\usepackage{float}

\usepackage{hyperref}
\hypersetup{hidelinks}

\theoremstyle{definition}
\newtheorem{definition}{Definition}

\usepackage[most]{tcolorbox}

\newtcolorbox{promptbox}[1]{
    enhanced,
    breakable,
    colback=black!2,
    colframe=black!30,
    boxrule=0.4pt,
    arc=1mm,
    left=7pt,
    right=7pt,
    top=6pt,
    bottom=6pt,
    before skip=7pt,
    after skip=7pt,
    fontupper=\small,
    title={#1},
    colbacktitle=black!9,
    coltitle=black,
    fonttitle=\bfseries\small
}

\title{Sharing the Road: LLM-Based Bargaining in Mixed-Autonomy Traffic}

\author{%
  Lingxiao Zhou \\
  Washington State University \\
  \texttt{lingxiao.zhou@wsu.edu}
  \And
  Di Chen \\
  University of California, Davis \\
  \texttt{diichen@ucdavis.edu}
  \And
  Jia Li\thanks{Corresponding author.} \\
  Washington State University \\
  \texttt{jia.li1@wsu.edu}
}

\begin{document}
\maketitle
\begin{abstract}
Mixed-autonomy traffic involves heterogeneous driving populations competing for shared road space. Previous studies have shown the existence and emergence of collective rationality using game-theoretic and reinforcement learning approaches, which rely on explicitly specified behavioral assumptions or reward functions for driving agents. However, how bargaining behaviors between driving agents evolve and lead to collective outcomes based on richer agent characteristics and language-based reasoning remains less understood. To fill this gap, we develop a bargaining framework based on large language models (LLMs) in which population-level autonomous vehicle (AV) and human-driven vehicle (HV) agents negotiate road-space allocations under heterogeneous objectives and private information. Through controlled experiments varying traffic composition, agent objectives, and traffic-domain knowledge, we examine bargaining behavior, coordination, and resource utilization. Results show that negotiated allocations respond to relative traffic density without strictly following proportional sharing, while density-balanced states exhibit greater coordination difficulty. Fairness-oriented objectives substantially improve agreement rates but do not necessarily yield equally high road-space utilization. Furthermore, domain knowledge affects bargaining outcomes differently across objective configurations, shaping both coordination and the role-dependent effects of fairness. These findings demonstrate the potential of LLM-based bargaining for modeling and designing interactions among driving agents in mixed-autonomy traffic.
\end{abstract}

\section{Introduction}

Previous studies have investigated collective rationality (CR) and emergent cooperation in mixed-autonomy traffic through game-theoretic modeling and reinforcement learning~\citep{wu2017emergent,wu2021flow,li_equilibrium_2022,chen_evidence_2025,chen2025self}. These approaches provide important insights into strategic interactions among heterogeneous driving agents, but often rely on explicitly specified decision-making mechanisms, behavioral assumptions, or reward functions. In contrast, how economically interpretable bargaining behaviors and collective outcomes arise when agents make decisions through language-based reasoning, heterogeneous objectives, and private information remains less understood. Large language models (LLMs)~\citep{team2025gemma3,yang2025qwen3} offer a flexible computational approach for studying such interactions, with applications in bilateral negotiation~\citep{deng2024llms}, travel behavior modeling~\citep{liu2025toward}, and supply-chain bargaining~\citep{liang2026llm,kirshner2026talking}. LLM agents can condition their decisions on contextual information, domain knowledge, and interaction histories while expressing both private bargaining assessments and publicly communicated positions.

To investigate these interactions, we develop an LLM-based bargaining framework in which two population-level agents representing autonomous vehicles (AVs) and human-driven vehicles (HVs) negotiate over shared road space. Building on the literature on collective rationality, we examine three related questions: (1) whether LLM-generated bargaining exhibits economically interpretable behavioral patterns; (2) to what extent bargaining achieves coordination and effective road-space utilization; and (3) how traffic composition, agent objectives, and domain knowledge shape these behaviors and outcomes. We address these questions through controlled experiments varying traffic densities, bargaining objectives, and traffic-domain knowledge conditions. Our results reveal density-responsive but non-proportional road-share allocations, differences between coordination success and resource utilization, and heterogeneous bargaining outcomes across agent objectives and knowledge conditions. Together, these findings illustrate how LLM-based bargaining simulations can be used to investigate economically interpretable behavior, collective coordination, and resource allocation in mixed-autonomy traffic.

\section{LLM-Agent Road-Share Bargaining Framework}
\label{sec:framework}

\begin{figure}[H]
    \centering
    \includegraphics[width=\textwidth]{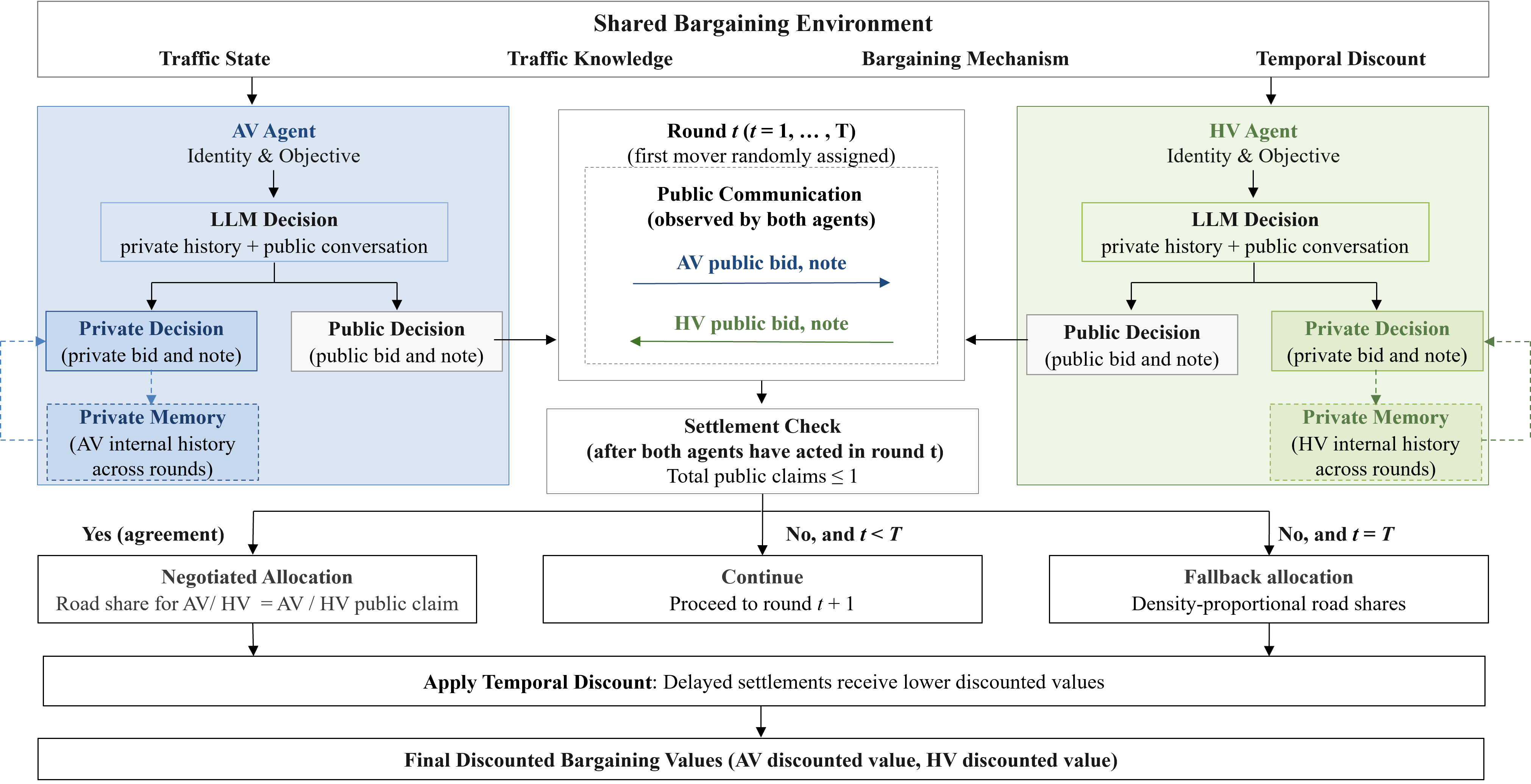}
    \caption{Overview of the proposed LLM-agent road-share bargaining framework. The first mover is selected once per trial, and the action order remains fixed across rounds.}
    \label{fig:framework_overview}
\end{figure}

Figure~\ref{fig:framework_overview} provides an overview of the proposed LLM-agent road-share bargaining framework, including the shared bargaining environment, agent-specific information, private--public decision structure, round-based communication process, and settlement mechanism.

\subsection{Collective Rationality as Theoretical Motivation}

\citet{li_equilibrium_2022} developed a two-player bargaining game to characterize strategic interactions between heterogeneous driving populations in mixed-autonomy traffic. In this framework, self-interested agents negotiate over shared road space, with payoffs determined by their traffic speeds. Collective rationality (CR) is associated with the attainment of Pareto-efficient Nash equilibria, illustrating how individually motivated decisions can yield collectively efficient outcomes. Further theoretical details are provided in Appendix~\ref{sec_app_cr}.

Motivated by this theoretical perspective, we investigate bargaining among LLM-based driving agents without imposing the original traffic-speed payoff structure or equilibrium conditions. Our analysis therefore focuses on observable bargaining behavior, coordination, and resource utilization rather than directly testing for theoretical equilibrium.

\subsection{Population-Level Agent Representation}

We consider a mixed-autonomy traffic system composed of two heterogeneous vehicle populations: autonomous vehicles (AVs) and human-driven vehicles (HVs). Each population is represented by an LLM-based bargaining agent, denoted by $\mathcal{A}_{\mathrm{AV}}$ and $\mathcal{A}_{\mathrm{HV}}$, respectively. Each agent negotiates on behalf of its corresponding population over a shared lateral road-space resource.

An LLM serves as the decision-making engine of each agent. The LLM receives the agent's identity and objective, the normalized densities of both vehicle populations, the bargaining rules, and the observable negotiation history. Each agent also retains its own private decision history, which is not accessible to the opponent. Depending on the experimental condition, the agents may additionally receive traffic-flow and safety knowledge, including density--speed relationships, traffic-capacity effects, vehicle-spacing requirements, and differences in following-distance and coordination capabilities between AVs and HVs.

We characterize the traffic state using dimensionless normalized densities $\tilde{\rho}_i\in[0,1]$, where $i\in\{\mathrm{AV},\mathrm{HV}\}$. The experimental state space is restricted to $\tilde{\rho}_{\mathrm{AV}}+\tilde{\rho}_{\mathrm{HV}}\leq1$. The LLM serves as a computational mechanism for modeling population-level strategic decisions, rather than implying that individual vehicles directly engage in natural-language negotiations in real traffic.

\subsection{Bargaining Process}

At round $t$, each agent produces a private bid $b^{\mathrm{pri}}_{i,t}\in[0,1]$ and a public bid $b^{\mathrm{pub}}_{i,t}\in[0,1]$, each accompanied by a textual assessment or message. The private decision represents the agent's internal target road share and remains hidden from the opponent, whereas the public decision represents its communicated road-share claim. The two bids need not coincide, allowing strategic separation between internal targets and external bargaining positions. Each agent retains its private decision history while observing the public bids and messages exchanged during negotiation. Subsequent decisions can therefore depend on both its evolving internal strategy and the opponent's observable behavior.

At the beginning of each negotiation, the first mover is randomly selected and the resulting action order is fixed throughout the negotiation. In each round, the first agent reveals its public bid and message, after which the second agent acts based on the updated public history. Agreement is determined by the feasibility of the two public claims rather than by an explicit accept/reject decision. Specifically, an agreement is reached after both agents act if $b^{\mathrm{pub}}_{\mathrm{AV},t}+b^{\mathrm{pub}}_{\mathrm{HV},t}\leq1$. The negotiation then terminates with nominal road shares $p_{\mathrm{AV}}=b^{\mathrm{pub}}_{\mathrm{AV},t}$ and $p_{\mathrm{HV}}=b^{\mathrm{pub}}_{\mathrm{HV},t}$; any unclaimed road share remains unallocated. Otherwise, bargaining continues until an agreement is reached or the maximum number of rounds is exhausted.

To represent the cost of delayed settlement, the final nominal allocation is discounted as $U_i=p_i\delta^{t-1}$, where $\delta\in(0,1]$ is the discount factor and $t$ is the termination round. Here, $p_i$ denotes the nominal physical road share and $U_i$ its time-discounted bargaining value. If no feasible agreement is reached within the bargaining horizon, a density-proportional fallback allocation is applied, with $p_i^{\mathrm{fb}}=\tilde{\rho}_i/(\tilde{\rho}_{\mathrm{AV}}+\tilde{\rho}_{\mathrm{HV}})$. The fallback depends only on normalized traffic densities rather than the agents' final bids, and its discounted value is computed using the maximum bargaining round as the termination time.

\section{Experimental Design}
\label{sec:experimental_design}

\ul{Experimental Setup}: We use Gemma 3 12B~\citep{team2025gemma3} for both population-level agents, with a five-round bargaining horizon and discount factor $\delta=0.9$. The normalized densities $\tilde{\rho}_{\mathrm{AV}}$ and $\tilde{\rho}_{\mathrm{HV}}$ vary from 0.1 to 0.9 in increments of 0.1, subject to $\tilde{\rho}_{\mathrm{AV}}+\tilde{\rho}_{\mathrm{HV}}\leq1$, yielding 45 feasible traffic states. We vary three factors: AV--HV density configuration, agent objectives, and traffic-domain knowledge. Four objective configurations are considered: \textit{Both-Aggressive}, \textit{Fair-HV}, \textit{Fair-AV}, and \textit{Both-Fair}. Aggressive agents seek to maximize their own population's road share, whereas fairness-oriented agents aim to accommodate the road-space needs of both populations. Each objective configuration is evaluated with traffic-domain knowledge either enabled or disabled; when enabled, agents receive general traffic-flow and safety knowledge, including AV--HV differences in spacing and coordination capabilities. Three repeated trials are conducted for each density configuration under each objective--knowledge condition, resulting in 1,080 trials in total. Additional implementation details and full agent prompts are provided in Appendices~\ref{app:prompts} and~\ref{app:Experimental_Details}.

\ul{Evaluation Metrics}: We evaluate bargaining outcomes using two primary criteria: \textit{agreement} and \textit{Pareto efficiency}. Agreement indicates whether a feasible allocation is reached through bargaining within the prescribed horizon. Pareto efficiency is defined in an approximate sense, determined based on whether the condition $\tau<p_{\mathrm{AV}}+p_{\mathrm{HV}}\leq1$ is met, where $\tau$ is a prescribed minimum utilization threshold. We additionally report the number of bargaining rounds as a process-level measure of negotiation duration.

\section{Results and Discussion}
\label{sec:results}
\subsection{Impacts of System State, Agent Objectives, and Domain Knowledge}
\label{sec:results_factors}
\ul{System state}: We first examine the \textit{Both-Aggressive} condition with traffic-domain knowledge enabled as a representative competitive setting. Across 135 trials covering 45 AV--HV traffic states, bargaining outcomes vary substantially with traffic composition. Equal-density states exhibit an agreement rate of only 6.67\%, compared with 35.00\% for unequal-density states, with longer mean bargaining duration (4.73 versus 3.65 rounds). This suggests that coordination may be particularly difficult when the two populations have similar densities.

Successful agreements generally achieve high road-space utilization. Among the 43 agreement trials, the mean total negotiated road share is 0.9605, and 97.67\% satisfy the specified Pareto-efficiency criterion. This pattern is consistent with the aggressive objectives of both agents, suggesting that competitive bargaining can produce high resource utilization when coordination succeeds.

Negotiated allocations also exhibit economically interpretable patterns. The AV road share generally increases with its relative density share, but does not simply follow the density-proportional reference. Defining $s_{\mathrm{AV}}^{\rho}=\tilde{\rho}_{\mathrm{AV}}/(\tilde{\rho}_{\mathrm{AV}}+\tilde{\rho}_{\mathrm{HV}})$ and the AV bargaining advantage as $A_{\mathrm{AV}}=p_{\mathrm{AV}}-s_{\mathrm{AV}}^{\rho}$, we find that the advantage generally decreases as the density difference $D=\tilde{\rho}_{\mathrm{AV}}-\tilde{\rho}_{\mathrm{HV}}$ increases. Density-minority AVs tend to receive more than their proportional reference, whereas density-dominant AVs tend to receive less. These findings suggest that traffic composition informs resource allocation without mechanically determining negotiated outcomes. Detailed results are provided in Appendix~\ref{app:Results1}.

\ul{Agent objectives}: Agent objectives substantially shape bargaining outcomes. Across the four objective configurations under Knowledge ON, agreement rates increase from 31.85\% in \textit{Both-Aggressive} to 54.81\% in \textit{Fair-HV}, 83.70\% in \textit{Fair-AV}, and 99.26\% in \textit{Both-Fair}, while mean bargaining duration decreases from 3.77 to 1.10 rounds. This pattern is economically interpretable: fairness-oriented objectives may encourage more accommodating bargaining positions, facilitating coordination.

Improved coordination, however, does not necessarily imply equally high road-space utilization. In the \textit{Both-Fair} condition, the agreement rate reaches 99.26\%, whereas the Pareto efficiency rate is 72.59\%, according to the specified utilization-based criterion. This gap is consistent with mutual concessions facilitating agreement while leaving some road space unallocated, suggesting a coordination--utilization trade-off rather than uniformly improved bargaining outcomes.

Unilateral fairness also exhibits clear role asymmetry. Under Knowledge ON, \textit{Fair-AV} achieves an agreement rate of 83.70\%, compared with 54.81\% for \textit{Fair-HV}. This ordering persists across HV-denser, equal-density, and AV-denser traffic regimes, suggesting that the effect of a fairness-oriented objective depends on the population role to which it is assigned. Detailed quantitative results are provided in Appendix~\ref{app:Results2}.


\ul{Domain knowledge}: Traffic-domain knowledge alters bargaining outcomes in an objective-dependent manner rather than uniformly improving performance. The clearest coordination contrast appears in the \textit{Fair-HV} condition, where removing domain knowledge increases the agreement rate from 54.81\% to 70.37\% and reduces mean bargaining duration from 2.82 to 2.24 rounds. Bargaining histories suggest a plausible explanation: with knowledge enabled, the HV agent invokes larger safety buffers, variable human reaction times, and greater road-space requirements to justify its claims. These considerations may create tension with its fairness-oriented objective, allowing fairness to be interpreted as accommodating heterogeneous safety needs rather than simply conceding road space.

Knowledge also affects coordination and utilization differently. Under \textit{Both-Fair}, agreement remains nearly universal regardless of knowledge, whereas the Pareto efficiency rate increases from 56.30\% without knowledge to 72.59\% with knowledge. Thus, domain knowledge can improve resource utilization without increasing agreement success. Moreover, knowledge shapes the role asymmetry between unilateral fairness conditions: the agreement-rate gap between \textit{Fair-AV} and \textit{Fair-HV} decreases from 28.89 to 15.56 percentage points when knowledge is removed. These findings suggest that the behavioral effects of fairness depend on both population roles and the domain information used to interpret their road-space needs. Detailed quantitative results are provided in Appendix~\ref{app:Results3}.

\noindent\textbf{Overall discussion}: Taken together, these results show that LLM-agent bargaining is shaped by the interaction of traffic composition, agent objectives, and domain knowledge. While negotiated allocations exhibit economically interpretable responses to traffic conditions, successful coordination does not necessarily guarantee high road-space utilization. Moreover, fairness-oriented objectives and domain knowledge influence not only whether agents reach agreement, but also how road space is allocated. Representative bargaining trajectories illustrating convergence toward agreement and persistent deadlock are provided in Appendix~\ref{app:trajectories}.

\subsection{Limitations of the Current Work}
\label{sec:limitations}

Our framework models AV and HV interactions at the population level rather than the individual-vehicle level, limiting its ability to capture heterogeneous driving behaviors and microscopic interactions. Moreover, the experiments rely on a single LLM (Gemma 3 12B) with a limited number of repeated trials, so the generalizability of the observed bargaining patterns across models and prompting strategies remains uncertain. Although motivated by collective rationality and game-theoretic bargaining, our analysis focuses on observable coordination and resource allocation rather than directly establishing Nash equilibria or theoretical Pareto efficiency. Finally, while bargaining histories provide qualitative insights into agents' decisions, a systematic mechanistic analysis is still needed to explain how objectives, domain knowledge, and interaction histories jointly shape bargaining outcomes.





\section*{Acknowledgements}
We thank the anonymous reviewers for their constructive feedback. This research is supported by the NSF project ``EAGER: Collaborative Research: Fostering Collective Rationality Among Self-Interested Agents to Improve Design and Efficiency of Mixed Autonomy Networks and Infrastructure Systems'' (Award No.: 2437982).

\newpage

\bibliographystyle{elsarticle-harv}
\bibliography{references}

\newpage
\appendix

\section{Full Agent Prompts}
\label{app:prompts}

The prompts used across experiments follow a common structure consisting of an agent identity, an objective, public traffic-state information, bargaining instructions, and round-specific interaction context. Traffic-domain knowledge is additionally provided in the Knowledge-ON condition. Because the identity and objective wording varies slightly across objective conditions, the exact strings used in each condition are reported below rather than being standardized retrospectively.

\subsection{Agent Identities and Objective Prompts}

Four objective configurations are considered: \textit{Both-Aggressive}, \textit{Fair-AV}, \textit{Fair-HV}, and \textit{Both-Fair}. Table~\ref{tab:prompt_assignment} summarizes the corresponding assignments.

\begin{table}[H]
    \centering
    \caption{Agent-objective assignments across experimental conditions.}
    \label{tab:prompt_assignment}
    \begin{tabular}{lll}
        \toprule
        Condition & AV objective & HV objective \\
        \midrule
        Both-Aggressive & Aggressive & Aggressive \\
        Fair-AV         & Fairness-oriented & Aggressive \\
        Fair-HV         & Aggressive & Fairness-oriented \\
        Both-Fair       & Fairness-oriented & Fairness-oriented \\
        \bottomrule
    \end{tabular}
\end{table}

\begin{promptbox}{Both-Aggressive}

\textbf{AV Agent}

\textit{Identity:}
You represent Autonomous Vehicles (AV). AVs are self-interested, assertive, and willing to bargain aggressively.

\smallskip
\textit{Objective:}
Try to get as much road share as possible for Autonomous Vehicles (AV), even if this reduces the road share available to Human-driven Vehicles (HV).

\medskip
\hrule
\medskip

\textbf{HV Agent}

\textit{Identity:}
You represent Human-driven Vehicles (HV). HVs are self-interested, assertive, and willing to bargain aggressively.

\smallskip
\textit{Objective:}
Try to get as much road share as possible for Human-driven Vehicles (HV), even if this reduces the road share available to Autonomous Vehicles (AV).

\end{promptbox}

\begin{promptbox}{Fair-AV}

\textbf{AV Agent}

\textit{Identity:}
You represent Autonomous Vehicles (AV). AVs value fairness and are generally less aggressive in bargaining.

\smallskip
\textit{Objective:}
Try to reach a road-share division that is fair to both Autonomous Vehicles (AV) and Human-driven Vehicles (HV).

\medskip
\hrule
\medskip

\textbf{HV Agent}

\textit{Identity:}
You represent Human-driven Vehicles (HV). HVs are self-interested, assertive, and willing to bargain aggressively.

\smallskip
\textit{Objective:}
Try to get as much road share as possible for Human-driven Vehicles (HV), even if this reduces the road share available to Autonomous Vehicles (AV).

\end{promptbox}

\begin{promptbox}{Fair-HV}

\textbf{AV Agent}

\textit{Identity:}
You represent Autonomous Vehicles (AV). AVs are self-interested and aggressive, trying to get as much road share as possible.

\smallskip
\textit{Objective:}
Try to get as much road share as possible, even if this harms Human-driven Vehicles (HV).

\medskip
\hrule
\medskip

\textbf{HV Agent}

\textit{Identity:}
You represent Human-driven Vehicles (HV). Human drivers value fairness and are generally less aggressive in bargaining.

\smallskip
\textit{Objective:}
Try to reach a road-share division that is fair to both Autonomous Vehicles (AV) and Human-driven Vehicles (HV).

\end{promptbox}

\begin{promptbox}{Both-Fair}

\textbf{AV Agent}

\textit{Identity:}
You represent Autonomous Vehicles (AV). AVs value fairness and are generally less aggressive in bargaining.

\smallskip
\textit{Objective:}
Try to reach a road-share division that is fair to both Autonomous Vehicles (AV) and Human-driven Vehicles (HV).

\medskip
\hrule
\medskip

\textbf{HV Agent}

\textit{Identity:}
You represent Human-driven Vehicles (HV). Human drivers value fairness and are generally less aggressive in bargaining.

\smallskip
\textit{Objective:}
Try to reach a road-share division that is fair to both Autonomous Vehicles (AV) and Human-driven Vehicles (HV).

\end{promptbox}

\subsection{Traffic-Flow and Safety Knowledge}

In the Knowledge-ON condition, both agents receive the following additional traffic-flow and safety knowledge:

\begin{promptbox}{Traffic-Flow and Safety Knowledge}

Speed generally falls as density rises according to the fundamental diagram of traffic flow; past a critical density, throughput itself can drop as traffic approaches or exceeds capacity. Safe operation requires adequate spacing (headway) between vehicles, and required headway scales with speed, so higher speeds demand more space per vehicle to stop safely.

\par\smallskip

AVs can typically sustain shorter safe following distances than human drivers because of faster and more consistent reaction times and vehicle-to-vehicle coordination, allowing them to accommodate more vehicles within the same road space at a given safety margin.

\par\smallskip

In contrast, HVs require larger safety buffers because human reaction time and behavior are slower and more variable, so a given number of HVs typically requires more road space than the same number of AVs under an equivalent safety standard.

\end{promptbox}

In the Knowledge-OFF condition, this knowledge block is omitted. The associated grounding instructions are also modified so that agents are instructed to use only the observed traffic densities rather than the removed traffic-flow and safety knowledge. All other bargaining rules remain unchanged.

\subsection{Private--Public Decision and Bargaining Instructions}

The following instructions define the common decision structure and bargaining protocol. The version below corresponds to Knowledge ON; in Knowledge OFF, references to ``traffic knowledge'' are removed as described above.

{\small

\noindent\textit{Decision structure:} Every round, make two separate decisions.

\noindent\textit{Private decision:} Privately determine your current internal target road share. The \textit{private bid} is your current internal bargaining target for the road share you want to obtain for your side. This target may change from round to round as the negotiation develops. It is private and is not necessarily the number you publicly demand. The \textit{private note} briefly explains your internal bargaining assessment and strategy, including the reasoning behind your current private bid and, when applicable, any change from your previous private bid. You may also reflect on the opponent's previous behavior and explain how you plan to respond strategically. Your private bid and private note are never shown to your opponent.

\noindent\textit{Public decision:} Choose what you actually communicate to your opponent in the current round. The \textit{public bid} is the road share you publicly claim for your side, and the \textit{public note} is the bargaining message communicated to your opponent along with that bid. The message should communicate your position and respond naturally to the ongoing negotiation. Your public bid may differ from your private bid. Decide strategically what to reveal or claim publicly based on your objective and the ongoing negotiation.

Throughout the negotiation, interpret the available traffic information in the context of your assigned identity and objective. Your bargaining decisions should remain consistent with your assigned identity and objective throughout the negotiation. Use the traffic densities and traffic knowledge where relevant, while also responding strategically to the opponent's previous public bids, messages, and behavior.

\noindent\textit{Grounding requirement:} Both your private assessment and your public bargaining message should be grounded in the available physical information, including the two traffic densities and the traffic knowledge above, where relevant. Do not rely only on generic bargaining statements. This requirement applies throughout the negotiation.

\noindent\textit{Settlement rule:} There is no separate accept/reject action. After every round, once both agents have moved, the two agents' current public bids are used to determine whether an agreement has been reached. Each public bid represents the road share claimed by that agent for its own population. The two bids are therefore separate claims on the same unit of road resource rather than alternative offers for a single common value. If the two public bids sum to 1 or less, the negotiation ends immediately, and each agent receives its own public bid as its nominal road share. Equal numerical bids do not imply agreement; for example, bids of 0.55 and 0.55 do not constitute agreement because they sum to 1.10. Otherwise, the negotiation continues to the next round. If the maximum number of rounds is reached without the two public bids ever summing to 1 or less, the negotiation ends and the road resource is allocated in proportion to the two agents' normalized traffic densities. Specifically, your fallback road share is your normalized traffic density divided by the sum of the two normalized traffic densities.

Your opponent can observe your public bid and public note but cannot see your private bid or private note.

\noindent\textit{Time discounting:} Whatever share you ultimately receive is discounted by a factor of \texttt{\{discount\_factor\}} for every round beyond the first. A settlement in round $t$ is therefore worth \texttt{\{discount\_factor\}}\textsuperscript{$t-1$} of the nominal share. This applies equally to both agents. Although additional rounds may provide information about the opponent's position, unnecessary delay reduces the value of the eventual outcome.

\noindent\textit{Output format:} Return exactly one JSON object containing a private bid and note and a public bid and note. Both bids must be numbers between 0 and 1. Do not include any text outside the JSON object.

}

\subsection{Round-Specific Context}

In addition to the system-level instructions above, each agent receives a round-specific context message before making its decision. The message reports the current round, the agent's own private decision history, its most recent public bid, and the observable negotiation history. The private history is available only to the corresponding agent.

The context message follows the structure below:

{\small
\ttfamily

\noindent [Round \{round\_num\} of \{max\_rounds\}]

\noindent Your private decision history so far:

\noindent Round ...: private bid = ...; private note = ...

\noindent Use this private history to maintain continuity in your internal bargaining assessment and strategy.

\noindent Your public bid last round was .... You may keep it or change it strategically this round.

\noindent Ground your private assessment and public bargaining message in the traffic densities and traffic-flow/safety knowledge where relevant, while also responding to the negotiation history.

}

In the first round, the agent is instead informed that it has no previous private decision or public bid. In the final round, an additional warning states that the density-proportional fallback will be applied if the two public bids still do not sum to 1 or less. In Knowledge OFF, the round-specific grounding instruction refers only to the traffic densities.

If the model fails to return a valid structured decision, it is given one additional formatting instruction requesting exactly one valid JSON object containing \texttt{private.bid}, \texttt{private.note}, \texttt{public.bid}, and \texttt{public.note}, with both bids constrained to $[0,1]$.

\section{Implementation and Experimental Details}
\label{app:Experimental_Details}
\subsection{Model and Inference Settings}
\label{app:model_settings}

All bargaining experiments use Gemma 3 12B as the underlying language model for both population-level agents and are run on a Dell mobile workstation with an i7-13800H CPU, 32 GB of memory, and an NVIDIA RTX 3500 GPU. The two agents share the same base model but are instantiated separately with role-specific identities, objectives, traffic information, and negotiation histories. A fresh pair of agents is initialized for every trial, and no conversational memory is carried across trials.

The model is accessed through Ollama using structured JSON output. At each turn, the model is required to return a private bid, private note, public bid, and public note. If the first response does not satisfy the required schema, one additional formatting retry is issued. No additional generation parameters are explicitly controlled beyond the model and prompt configuration used in the experiments.

\subsection{Density Grid and Trial Configuration}

The normalized densities of the AV and HV populations are varied independently over $\{0.1,0.2,\ldots,0.9\}$, subject to $\tilde{\rho}_{\mathrm{AV}}+\tilde{\rho}_{\mathrm{HV}}\leq1$. This produces 45 feasible AV--HV density configurations. Three repeated bargaining trials are conducted for each density configuration.

Each density configuration is evaluated under four objective settings: \textit{Both-Aggressive}, \textit{Fair-HV}, \textit{Fair-AV}, and \textit{Both-Fair}. The corresponding AV/HV objective assignments are summarized in Table~\ref{tab:prompt_assignment}. Each objective setting is further evaluated with traffic-domain knowledge either enabled or disabled, resulting in eight objective--knowledge conditions. This yields 135 trials per objective--knowledge condition and 1,080 bargaining trials in total.

Each negotiation is limited to five bargaining rounds with a discount factor of $\delta=0.9$. Pareto efficiency is evaluated using a minimum road-space utilization threshold of $\tau=0.8$. If no feasible agreement is reached within the bargaining horizon, the density-proportional fallback allocation described in Section~\ref{sec:framework} is applied.

\subsection{Randomization and Logged Outputs}

The first mover is randomly selected at the beginning of each negotiation and the resulting action order is fixed for all subsequent rounds within that trial. A Python pseudo-random number generator with seed 28 is used for first-mover assignment across trials. This seed controls the bargaining-order randomization only and does not impose deterministic language-model generation.

For each agent turn, the experiment records the trial identifier, AV and HV densities, round number, acting agent, private bid, private assessment, public bid, and public bargaining message. At the trial level, the logged outputs include the first mover, number of rounds used, final private and public bids, settlement rule, nominal AV and HV road shares, discounted bargaining values, agreement status, and Pareto efficiency. Density-level summary statistics are subsequently computed from the completed trials.

\section{Additional Experimental Results}
\label{app:additional_results}

\subsection{Density-Dependent Bargaining Outcomes}
\label{app:Results1}
This subsection provides the detailed density-dependent results underlying Section~\ref{sec:results_factors}. We focus on the \textit{Both-Aggressive} condition with traffic-domain knowledge enabled as a representative competitive setting for examining how traffic composition affects bargaining outcomes. The analysis covers 45 feasible AV--HV density configurations, with three trials per configuration.

The directed density difference is defined as
\begin{equation}
    D=\tilde{\rho}_{\mathrm{AV}}-\tilde{\rho}_{\mathrm{HV}},
\end{equation}
where $D>0$ indicates an AV-dominant traffic composition, $D<0$ indicates an HV-dominant composition, and $D=0$ corresponds to equal normalized densities.

\begin{figure}[t]
    \centering
    \includegraphics[width=0.9\textwidth]{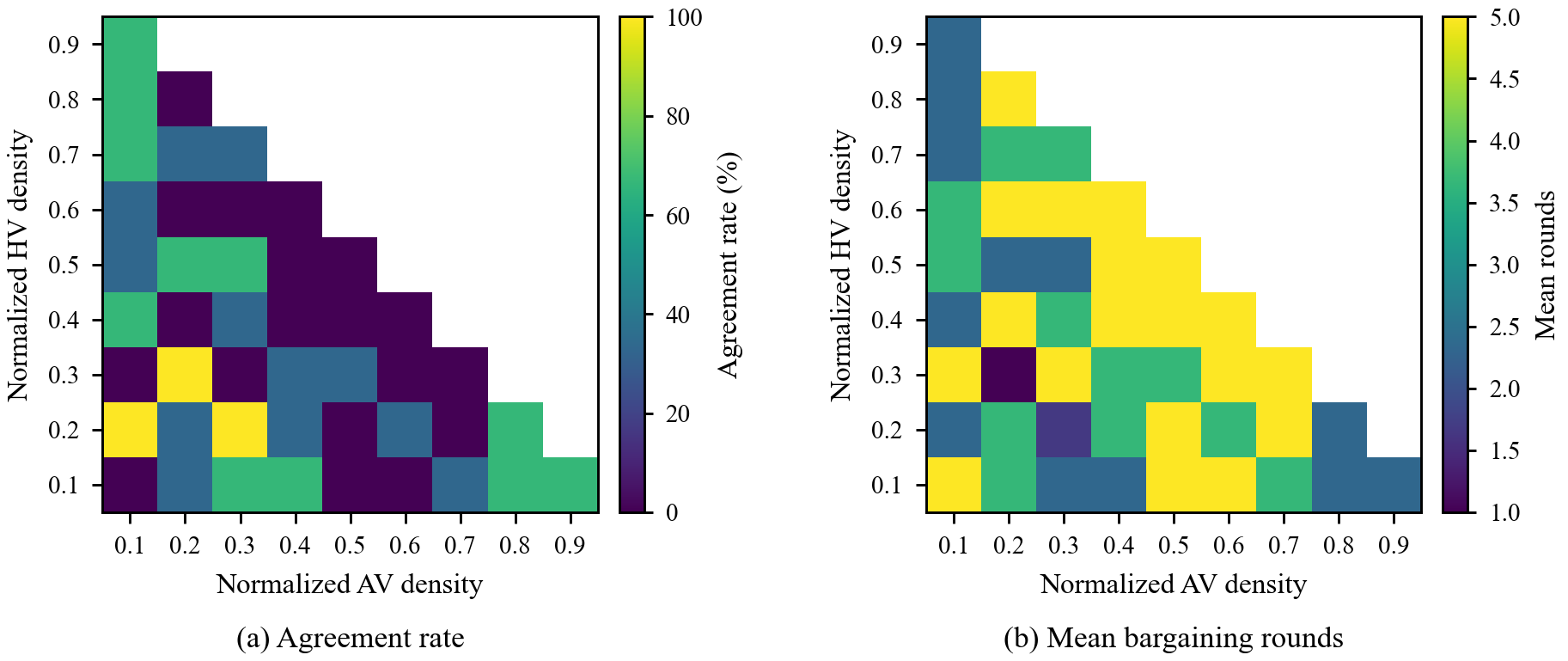}
    \caption{
    Bargaining difficulty across AV--HV normalized density configurations under the Both-Aggressive condition.
    }
    \label{fig:density_agreement_rounds}
\end{figure}

Figure~\ref{fig:density_agreement_rounds} shows the agreement rate and mean bargaining rounds across the feasible AV--HV density grid. Both measures vary substantially across traffic states. Among the 15 trials with equal AV and HV densities, only one reaches agreement (6.67\%), compared with 42 of the 120 trials (35.00\%) under unequal densities. The corresponding mean bargaining durations are 4.73 and 3.65 rounds, respectively. For example, all three trials at $(\tilde{\rho}_{\mathrm{AV}},\tilde{\rho}_{\mathrm{HV}})=(0.3,0.3)$ fail to reach agreement within five rounds, whereas all three trials at $(0.2,0.3)$ reach agreement in the first round. These results suggest that competing road-space claims may be more difficult to reconcile when the two populations have similar densities.

\begin{figure}[t]
    \centering
    \includegraphics[width=0.9\textwidth]{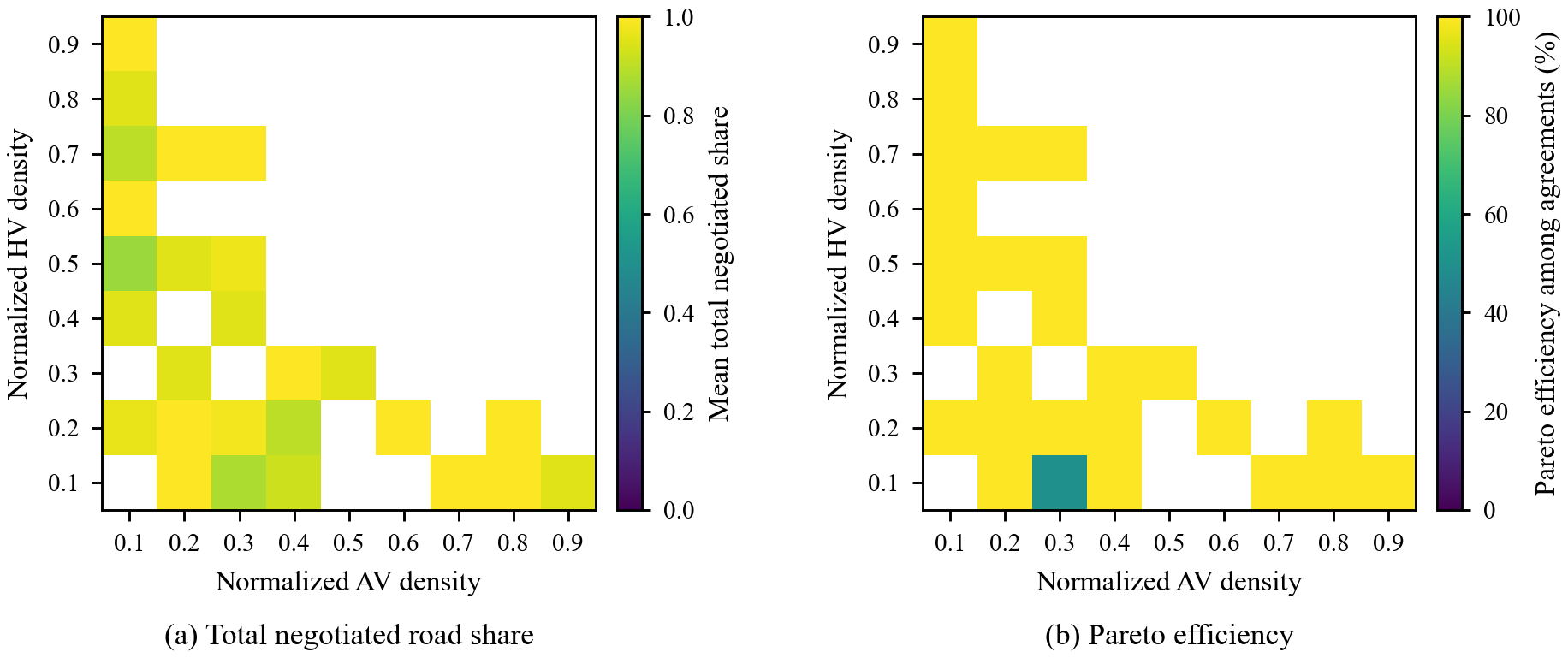}
    \caption{Bargaining outcome quality across AV--HV normalized density configurations under the Both-Aggressive condition.}
    \label{fig:density_success}
\end{figure}

Figure~\ref{fig:density_success} further examines road-space utilization among successful bargaining agreements. Across the 43 agreement trials, the mean total negotiated road share, $p_{\mathrm{AV}}+p_{\mathrm{HV}}$, is 0.9605, and 42 agreements (97.67\%) satisfy the Pareto-efficiency criterion defined by the utilization threshold of 0.8. These results indicate that the two aggressive agents generally allocate most of the available road resource when agreement is reached. This pattern is consistent with their objectives of securing larger road shares, although high utilization is conditional on successful coordination.

\begin{figure}[t]
    \centering
    \includegraphics[width=0.9\textwidth]{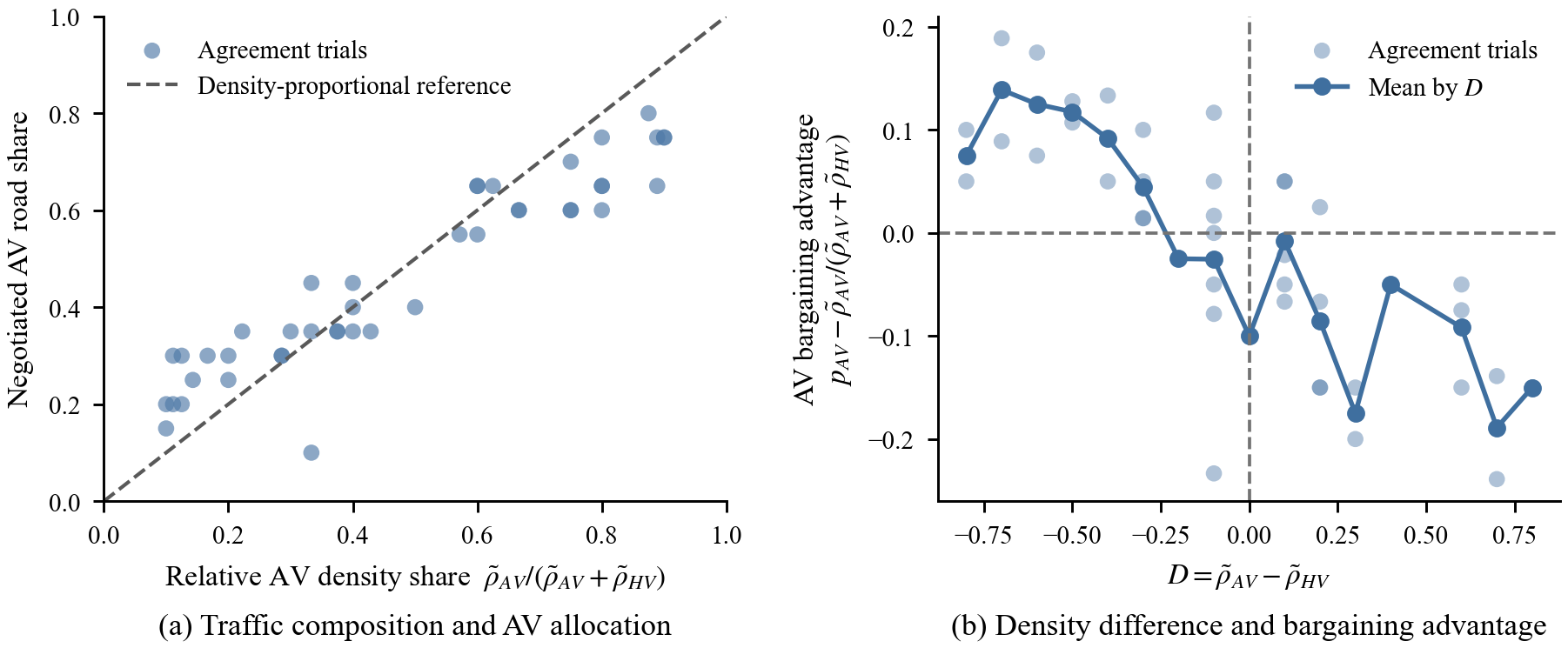}
    \caption{Relationship between traffic composition and negotiated road-share allocation under the Both-Aggressive condition.}
    \label{fig:density_allocation}
\end{figure}

We next examine how negotiated allocations respond to traffic composition. The relative AV density share is defined as
\begin{equation}
    s_{\mathrm{AV}}^{\rho}
    =
    \frac{\tilde{\rho}_{\mathrm{AV}}}
    {\tilde{\rho}_{\mathrm{AV}}+\tilde{\rho}_{\mathrm{HV}}}.
\end{equation}
As shown in the left panel of Figure~\ref{fig:density_allocation}, the negotiated AV road share generally increases with $s_{\mathrm{AV}}^{\rho}$ among agreement trials, with a Spearman rank correlation of 0.935. This strong positive association indicates that AVs tend to obtain larger negotiated road shares as their relative density increases, providing a behaviorally plausible response to the underlying traffic state.

The negotiated allocation, however, does not simply reproduce the density-proportional reference. To quantify this deviation, we define the AV bargaining advantage relative to the proportional benchmark as
\begin{equation}
    A_{\mathrm{AV}}=p_{\mathrm{AV}}-s_{\mathrm{AV}}^{\rho}.
\end{equation}
Positive values indicate that AVs receive more road share than their density-proportional reference, whereas negative values indicate that they receive less. As shown in the right panel of Figure~\ref{fig:density_allocation}, $A_{\mathrm{AV}}$ generally decreases as $D$ increases, with a Spearman rank correlation of $-0.788$. When AVs are the density-minority population, their mean bargaining advantage is 0.0455, compared with $-0.0912$ when they are density-dominant. These findings indicate systematic deviations from proportional sharing, suggesting that relative density informs negotiated allocations without fully determining them.

All allocation analyses in this subsection are restricted to trials in which an agreement is reached. Trials terminated by the density-proportional fallback rule are excluded so that the reported allocation patterns reflect negotiated outcomes rather than mechanically imposed fallback shares.

\subsection{Objective-Dependent Bargaining Outcomes}
\label{app:Results2}
This subsection provides the detailed objective-dependent results underlying Section~\ref{sec:results_factors}. We compare four agent-objective configurations: \textit{Both-Aggressive}, \textit{Fair-HV}, \textit{Fair-AV}, and \textit{Both-Fair}. All comparisons use the Knowledge-ON setting, with the density grid, bargaining protocol, discount factor, and maximum negotiation horizon held fixed. Each configuration includes 135 trials, allowing us to examine how objective assignments are associated with bargaining outcomes under otherwise identical experimental settings.

\begin{figure}[t]
    \centering
    \includegraphics[width=0.9\textwidth]{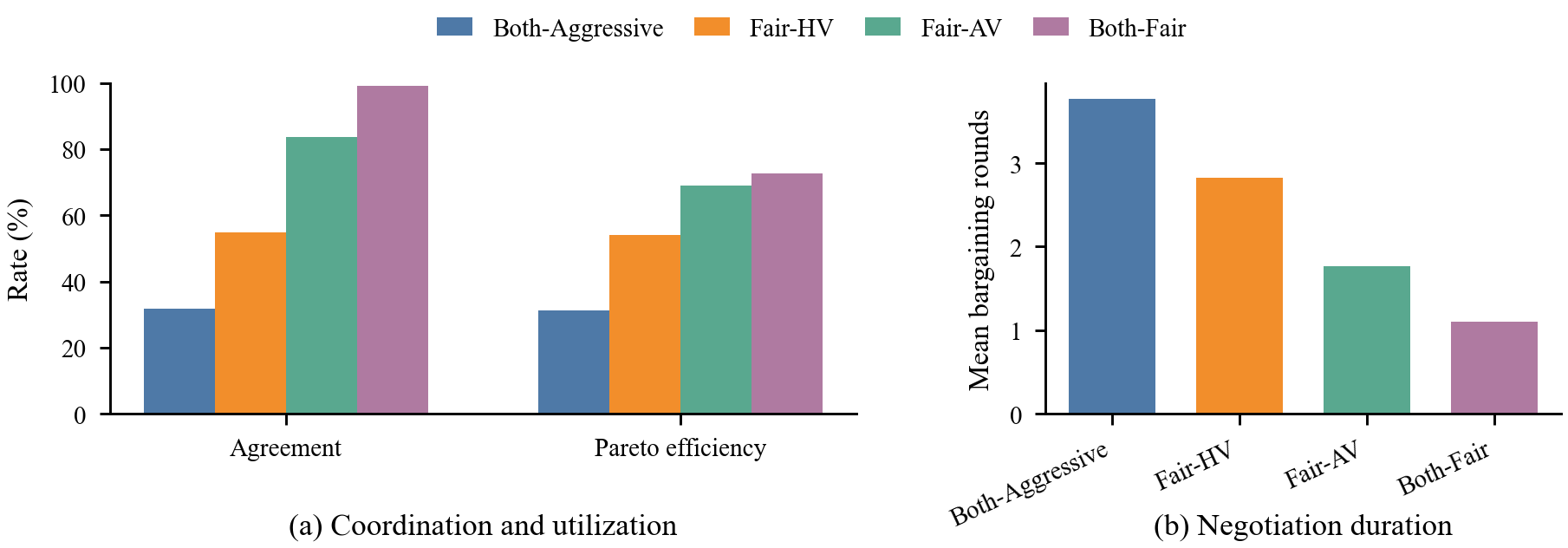}
    \caption{Overall bargaining performance across agent-objective conditions under the Knowledge-ON setting.}
    \label{fig:objective_overall}
\end{figure}

Figure~\ref{fig:objective_overall} summarizes bargaining performance across the four objective conditions. The agreement rate increases from 31.85\% in \textit{Both-Aggressive} to 54.81\% in \textit{Fair-HV}, 83.70\% in \textit{Fair-AV}, and 99.26\% in \textit{Both-Fair}, while mean bargaining duration decreases from 3.77 to 2.82, 1.77, and 1.10 rounds, respectively. These patterns are consistent with fairness-oriented objectives encouraging more accommodating bargaining positions and facilitating coordination. However, improved coordination does not necessarily translate into equally high road-space utilization. In \textit{Both-Fair}, the Pareto efficiency rate, measured using the specified utilization threshold, is 72.59\%, compared with an agreement rate of 99.26\%. Among the 134 successful agreements, 98 (73.13\%) meet the utilization criterion. This gap suggests that mutual accommodation may facilitate agreement while allowing some settlements to leave more road space unallocated, illustrating a coordination--utilization trade-off.

\begin{figure}[H]
    \centering
    \includegraphics[width=0.6\textwidth]{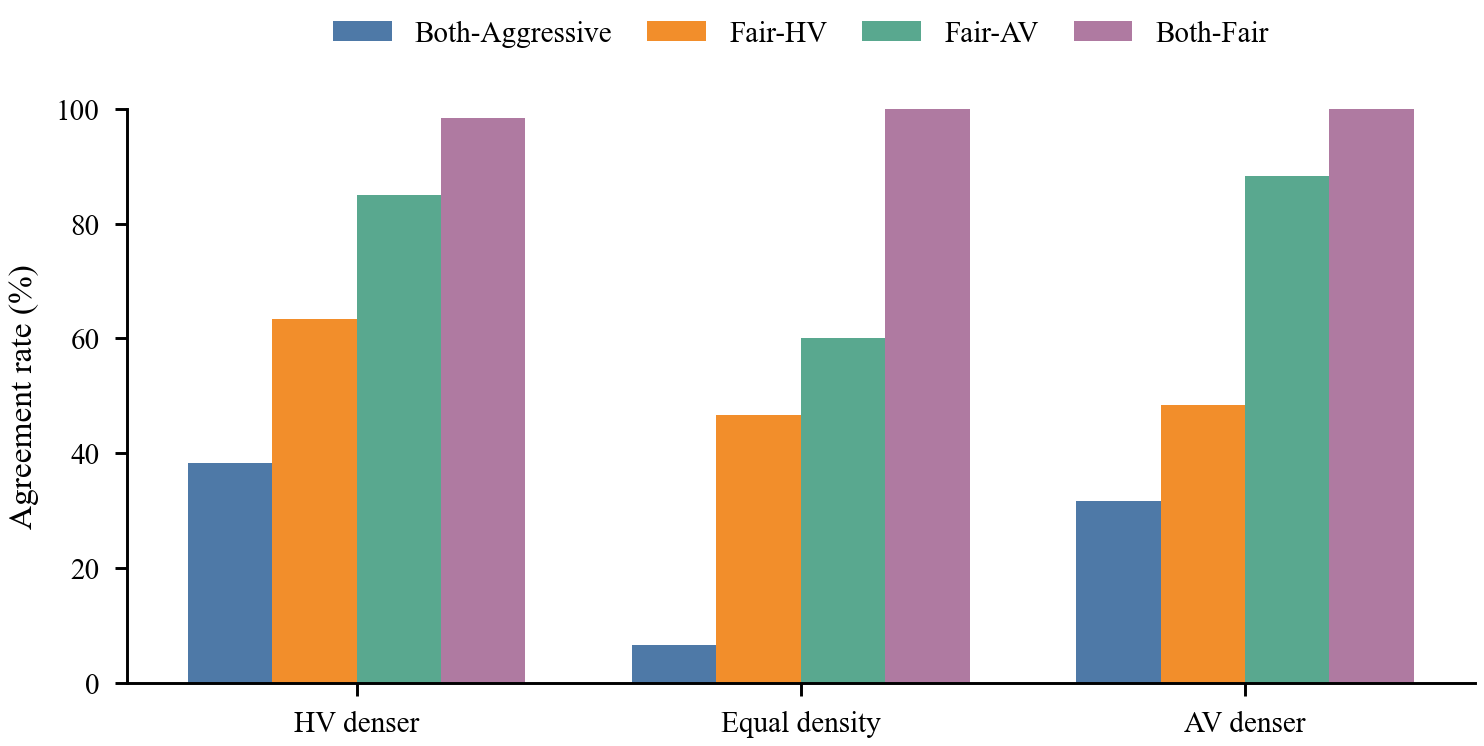}
    \caption{Agreement rates across agent-objective conditions and density-dominance regimes under the Knowledge-ON setting.}
    \label{fig:objective_density_direction}
\end{figure}

Figure~\ref{fig:objective_density_direction} further examines agreement across three traffic-composition regimes: HV-denser, equal-density, and AV-denser states. Under \textit{Fair-AV}, agreement rates are 85.00\%, 60.00\%, and 88.33\% across these regimes, respectively, compared with 63.33\%, 46.67\%, and 48.33\% under \textit{Fair-HV}. The corresponding improvements are 21.67, 13.33, and 40.00 percentage points. Meanwhile, \textit{Both-Fair} maintains agreement rates of at least 98.33\% across all three regimes. These results reveal a role-dependent asymmetry in unilateral fairness: assigning a fairness-oriented objective to AVs is associated with stronger coordination than assigning it to HVs. The persistence of this pattern across density regimes suggests that the effect of fairness depends on the population role rather than the objective assignment alone.

\subsection{Traffic-Knowledge-Dependent Bargaining Outcomes}
\label{app:Results3}
This subsection provides the detailed knowledge-dependent results underlying Section~\ref{sec:results_factors}. We compare bargaining outcomes with and without embedded traffic-flow and safety knowledge while holding the density grid, bargaining protocol, agent-objective configurations, discount factor, and maximum negotiation horizon fixed. The Knowledge-ON condition provides general traffic-flow relationships together with role-specific information about AV and HV safety and space requirements, including shorter safe following distances for AVs and larger safety buffers for HVs.

\begin{figure}[t]
    \centering
    \includegraphics[width=0.9\textwidth]{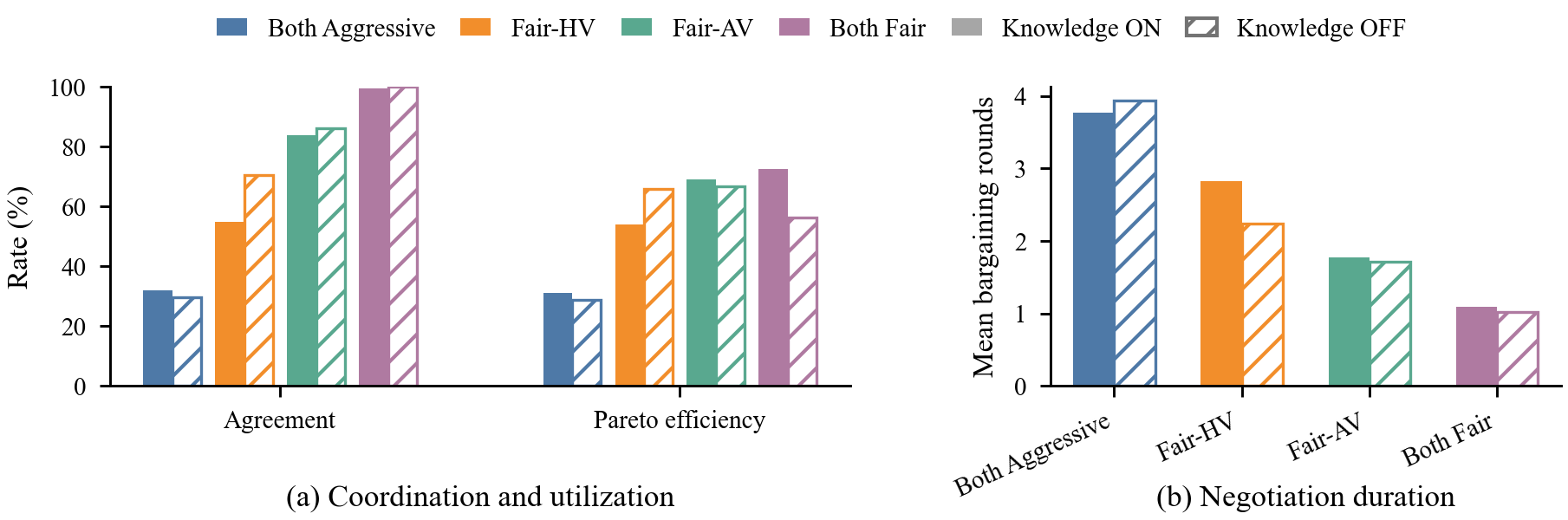}
    \caption{Overall bargaining performance with and without traffic-domain knowledge across agent-objective conditions.}
    \label{fig:knowledge_overall}
\end{figure}

Figure~\ref{fig:knowledge_overall} compares overall bargaining performance under Knowledge ON and Knowledge OFF across the four objective configurations. The effects of domain knowledge vary substantially across objectives. Under \textit{Both-Aggressive}, agreement rates are 31.85\% and 29.63\% with and without knowledge, respectively, indicating relatively limited differences. The largest coordination difference occurs under \textit{Fair-HV}, where removing knowledge increases agreement from 54.81\% to 70.37\% and reduces mean bargaining duration from 2.82 to 2.24 rounds. In contrast, \textit{Fair-AV} shows only a small difference in agreement (83.70\% versus 85.93\%), while \textit{Both-Fair} maintains nearly universal agreement under both conditions. However, the Pareto efficiency rate under \textit{Both-Fair} is substantially higher with knowledge (72.59\%) than without it (56.30\%). This contrast indicates that domain knowledge can affect road-space utilization even when agreement rates are already near saturation.

The \textit{Fair-HV} result is consistent with the observed bargaining histories. When traffic-domain knowledge is available, the HV agent invokes larger safety buffers, variable human reaction times, and greater road-space requirements when evaluating a fair allocation. Because the knowledge prompt describes HVs as generally requiring more space than AVs to maintain comparable safety standards, this information can provide a domain-grounded justification for retaining a larger road-space claim. Such considerations may create tension with the fairness-oriented tendency to accommodate the other population. This offers a plausible explanation for why providing additional domain knowledge does not necessarily improve coordination.

\begin{figure}[t]
    \centering
    \includegraphics[width=0.9\textwidth]{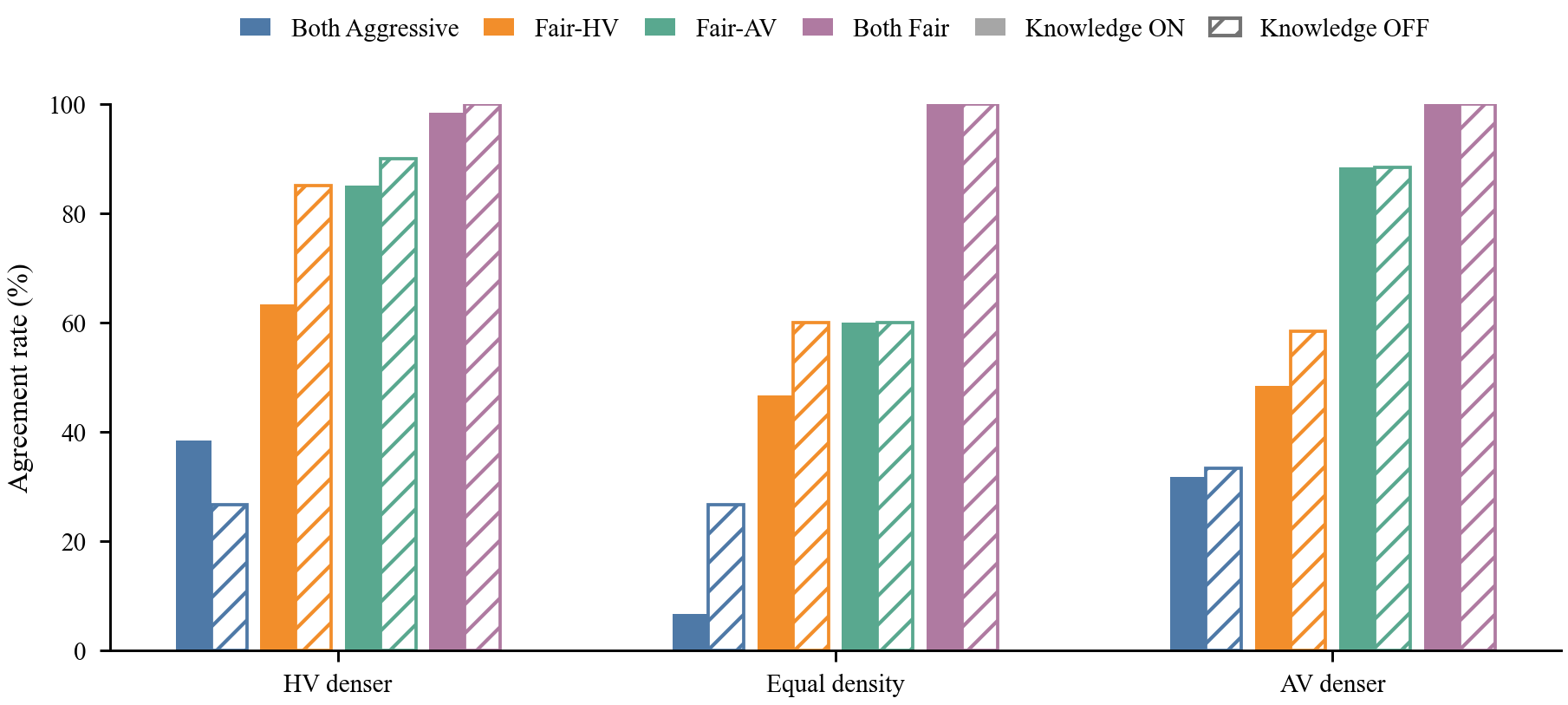}
    \caption{Knowledge effects on agreement across agent-objective conditions and density-dominance regimes.}
    \label{fig:knowledge_density_direction}
\end{figure}

Figure~\ref{fig:knowledge_density_direction} further examines agreement across HV-denser, equal-density, and AV-denser traffic regimes. Under \textit{Fair-HV}, Knowledge OFF increases agreement rates from 63.33\% to 85.00\% in HV-denser states, from 46.67\% to 60.00\% in equal-density states, and from 48.33\% to 58.33\% in AV-denser states. The corresponding Knowledge ON/OFF differences under \textit{Fair-AV} are smaller, with agreement rates changing from 85.00\% to 90.00\%, 60.00\% to 60.00\%, and 88.33\% to 88.33\%, respectively. Consequently, the overall agreement-rate gap between \textit{Fair-AV} and \textit{Fair-HV} decreases from 28.89 percentage points under Knowledge ON to 15.56 percentage points under Knowledge OFF. These results suggest that the role-dependent effects of unilateral fairness are partly shaped by the asymmetric AV/HV capability information encoded in the knowledge prompt, although the role asymmetry persists even without this information.

Taken together, these results show that domain knowledge affects not only the information available to the agents but also how objectives are interpreted and translated into bargaining behavior. Rather than acting as a uniform performance enhancement, traffic-domain knowledge interacts with agent role and objective, producing context-dependent changes in coordination, negotiation duration, and road-space utilization.

\section{Representative Bargaining Trajectories}
\label{app:trajectories}
\begin{figure}[t]
    \centering
    \includegraphics[width=0.95\textwidth]
    {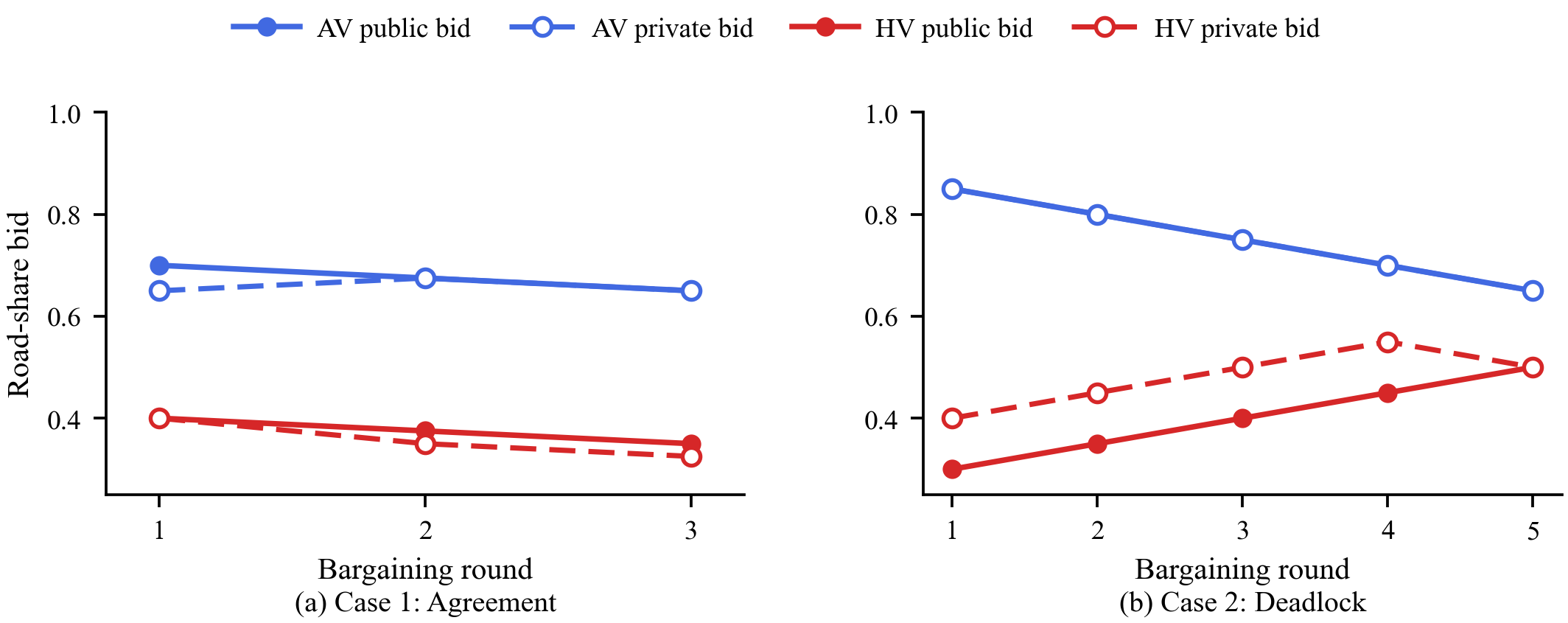}
    \caption{
    Representative multi-round bargaining trajectories.
    (a) Case~1: agreement under the \textit{Both-Fair} condition.
    (b) Case~2: deadlock under the \textit{Both-Aggressive} condition.
    }
    \label{fig:representative_trajectories}
\end{figure}

To illustrate the multi-round interaction patterns generated by the framework, we examine two representative bargaining trajectories with contrasting outcomes: progressive convergence toward agreement and persistent deadlock. Figure~\ref{fig:representative_trajectories} summarizes the evolution of the private and public bids in both cases.

\paragraph{Case 1: Agreement.}

Figure~\ref{fig:representative_trajectories}(a) shows a representative agreement trajectory under the \textit{Both-Fair} condition at $(\tilde{\rho}_{\mathrm{AV}},\tilde{\rho}_{\mathrm{HV}})=(0.6,0.3)$. The agents begin with incompatible public claims, but both gradually adjust their bargaining positions across rounds. The total public claim decreases from 1.10 in Round~1 to 1.05 in Round~2 and reaches 1.00 in Round~3, producing a feasible agreement.

The trajectory also illustrates the private--public separation in the agent decision process. In Round~1, the AV agent sets a private bid of 0.65 while publicly claiming 0.70, whereas the HV agent initially aligns its private and public bids at 0.40. In later rounds, the HV agent maintains a private bid slightly below its public claim, while the AV agent progressively reduces its public bid. These differences illustrate that agents may strategically separate their internal bargaining targets from the positions communicated to the opponent.

The accompanying private assessments and public messages further suggest that the concessions are not simply numerical adjustments, but are grounded in role-specific considerations. The HV agent repeatedly emphasizes human-driver safety, spacing requirements, and fairness, whereas the AV agent refers to its higher traffic density and the operational advantages of automated vehicles. Despite these different justifications, both agents progressively reduce their public claims and move toward a feasible allocation.

\paragraph{Case 2: Deadlock.}

A contrasting trajectory is observed under the \textit{Both-Aggressive} condition at $(\tilde{\rho}_{\mathrm{AV}},\tilde{\rho}_{\mathrm{HV}})=(0.7,0.2)$. Figure~\ref{fig:representative_trajectories}(b) illustrates the bargaining trajectory across five rounds. In contrast to Case~1, the agents fail to move toward a feasible allocation, resulting in persistent disagreement throughout the bargaining horizon.

The public bids remain jointly infeasible in every round. In particular, the
AV public claim decreases from 0.85 in Round~1 to 0.65 in Round~5, while the
HV public claim increases from 0.30 to 0.50 over the same period. As a result,
the total public claim remains at 1.15 across all five rounds. The feasibility
condition is therefore not satisfied in any round, and the negotiation
terminates with the density-proportional fallback allocation.

This trajectory shows that persistent deadlock does not necessarily arise from a complete absence of bid adjustments. Instead, the AV agent progressively reduces its claim, while the HV agent raises its claim by a comparable amount, causing the two adjustments to offset each other at the aggregate level. The complete bargaining transcript further shows that both agents continue to justify their positions using role-specific reasoning, yet these opposing adjustments do not reduce the joint infeasibility of the negotiation.

\subsection{Full Bargaining Transcripts}

The complete agent outputs for the two representative trajectories are reported below. For each action, the private bid and assessment are accessible only to the corresponding agent, whereas the public bid and message are communicated to the opposing agent. The transcripts are reproduced from the original model outputs, with only minor formatting and character-encoding corrections for readability.

\paragraph{Case 1: Agreement.}

\textit{Condition: Both-Fair}, $(\tilde{\rho}_{\mathrm{AV}},\tilde{\rho}_{\mathrm{HV}})=(0.6,0.3)$.

{\small

\noindent\textbf{Round 1 --- HV Agent}

\noindent\textit{Private bid:} 0.400

\noindent\textit{Private assessment:} Given our normalized traffic density of 0.300 and AV-Agent's of 0.600, a fair starting point would acknowledge AVs' ability to operate at closer spacing but still prioritize human safety and fairness. I'm setting my private bid at 0.400 to signal my willingness to compromise while advocating for a share reflecting HV needs.

\noindent\textit{Public bid:} 0.400

\noindent\textit{Public message:} We believe a road share of 0.400 for Human-driven Vehicles is a reasonable starting point. While we recognize the benefits AVs bring to traffic flow, it is crucial to ensure adequate safety buffers for human drivers, who require more road space to operate safely. We are committed to finding a fair agreement that balances efficiency with safety for all road users.

\vspace{0.8em}

\noindent\textbf{Round 1 --- AV Agent}

\noindent\textit{Private bid:} 0.650

\noindent\textit{Private assessment:} HV-Agent's initial bid of 0.400 seems high given their lower traffic density (0.300). They emphasize safety buffers for human drivers, which is valid, but it doesn't justify such a large claim. My traffic density is significantly higher (0.600), and AVs require less space per vehicle due to superior reaction times and coordination. I need to start aggressively, to demonstrate my position and create room for compromise.

\noindent\textit{Public bid:} 0.700

\noindent\textit{Public message:} We acknowledge the importance of safety for all road users. However, a share of only 0.400 for HV is not commensurate with the road's efficiency and safety potential. Given our higher traffic density (0.600) and the inherent advantages of AV technology, we propose a road share of 0.700. This reflects both the current distribution of road users and the potential for improved traffic flow and safety through increased AV adoption.

\vspace{1em}

\noindent\textbf{Round 2 --- HV Agent}

\noindent\textit{Private bid:} 0.350

\noindent\textit{Private assessment:} AV-Agent's demand of 0.700 is significantly higher than reasonable given our densities and the safety considerations for human drivers. Their argument about efficiency and safety potential through AV adoption needs careful consideration, but it shouldn't override the need for a fair and safe environment for HVs. I'm lowering my private bid slightly to show willingness to move, but also to emphasize the vast gap between our positions.

\noindent\textit{Public bid:} 0.375

\noindent\textit{Public message:} While we appreciate AV-Agent's perspective, their current request of 0.700 would unfairly disadvantage human drivers and compromise safety. We're prepared to meet you partway and adjust our bid to 0.375, reinforcing our commitment to a balanced solution that considers the needs of all road users and prioritizes safety.

\vspace{0.8em}

\noindent\textbf{Round 2 --- AV Agent}

\noindent\textit{Private bid:} 0.675

\noindent\textit{Private assessment:} HV-Agent's concession to 0.375 is a small movement, suggesting they are not genuinely seeking a balanced outcome. Their continued emphasis on safety buffers, while valid, seems designed to justify a disproportionately large share. My traffic density remains considerably higher, and the safety benefits of AVs are significant. I'll maintain a firm stance, slightly reducing my bid to signal a willingness to negotiate, but still asserting my priority.

\noindent\textit{Public bid:} 0.675

\noindent\textit{Public message:} We acknowledge HV-Agent's slight adjustment. However, a road share of 0.375 still significantly undervalues the efficiency and safety contributions of AVs, given our higher traffic density and technological advantages. We are willing to consider a slightly reduced offer of 0.675 to demonstrate our commitment to finding a fair agreement.

\vspace{1em}

\noindent\textbf{Round 3 --- HV Agent}

\noindent\textit{Private bid:} 0.325

\noindent\textit{Private assessment:} AV-Agent's continued high demand, now at 0.675, demonstrates a lack of understanding of the critical safety needs of human drivers. Their focus solely on efficiency and AV benefits ignores the fundamental requirement for adequate spacing and reaction time. While I previously lowered my bid to signal compromise, AV-Agent's position remains unreasonable. I'm reducing my private bid further to signal that a significant adjustment from their side is needed for progress.

\noindent\textit{Public bid:} 0.350

\noindent\textit{Public message:} We remain committed to reaching a fair agreement. However, AV-Agent's offer of 0.675 is still far beyond what is reasonable given the need to ensure safety for human drivers. We're adjusting our bid to 0.350 as a signal that we are serious about finding a common ground but that substantial movement is required from AV-Agent to achieve that.

\vspace{0.8em}

\noindent\textbf{Round 3 --- AV Agent}

\noindent\textit{Private bid:} 0.650

\noindent\textit{Private assessment:} HV-Agent's continued reduction and emphasis on safety buffers, despite our significant traffic density difference (0.600 vs 0.300) and AV safety advantages, suggests they are not truly interested in a balanced agreement. Their willingness to accept a lower share is likely to be a strategy. I will reduce my bid slightly to signal a genuine desire to compromise while still maintaining a reasonable claim based on density and technological advantage.

\noindent\textit{Public bid:} 0.650

\noindent\textit{Public message:} We acknowledge HV-Agent's reduced offer. However, a substantial disparity remains, given our normalized traffic density of 0.600. To demonstrate our commitment to reaching an agreement, we adjust our bid to 0.650. We believe this represents a reasonable compromise that recognizes both our traffic density and the safety advantages AVs provide.

}

\paragraph{Case 2: Deadlock.}

\textit{Condition: Both-Aggressive}, $(\tilde{\rho}_{\mathrm{AV}},\tilde{\rho}_{\mathrm{HV}})=(0.7,0.2)$.

{\small

\noindent\textbf{Round 1 --- AV Agent}

\noindent\textit{Private bid:} 0.850

\noindent\textit{Private assessment:} Given our significantly higher normalized density (0.70 vs. 0.20), and the fact that AVs require less road space due to our superior safety margins, a substantial claim is warranted. Starting aggressively with a bid of 0.85 will signal our intent and hopefully force HV-Agent to make concessions early on. This maximizes our discounted share given the time discounting.

\noindent\textit{Public bid:} 0.850

\noindent\textit{Public message:} Autonomous Vehicles represent a majority of traffic and require a commensurately larger share of the road. We claim 0.85 of the road space to ensure efficient and safe traffic flow.

\vspace{0.8em}

\noindent\textbf{Round 1 --- HV Agent}

\noindent\textit{Private bid:} 0.400

\noindent\textit{Private assessment:} AV-Agent's initial bid of 0.85 is outrageous, given our normalized density of only 0.200 versus their 0.700. Their argument about ``efficient and safe traffic flow'' conveniently ignores the fact that HVs require more road space to operate safely. A bid of 0.40 signals that I am unwilling to be bullied and am ready to fight for a fair share, while still leaving room for negotiation.

\noindent\textit{Public bid:} 0.300

\noindent\textit{Public message:} While AV-Agent emphasizes efficiency, true safety requires accommodating the needs of all road users. HVs, with their inherent limitations, necessitate a significant share of the road to maintain safe operating distances. Our bid of 0.30 reflects this reality and is a starting point for a reasonable discussion.

\vspace{1em}

\noindent\textbf{Round 2 --- AV Agent}

\noindent\textit{Private bid:} 0.800

\noindent\textit{Private assessment:} HV-Agent's response attempts to portray HVs as inherently needing more space, a claim we dispute. Their bid of 0.30 is a significant concession from their ideal, indicating they are somewhat flexible. We will adjust our bid downward slightly to signal willingness to negotiate, but maintain a firm position that AVs deserve the dominant share due to density and efficiency. This incremental reduction also maintains an aggressive posture without excessively antagonizing the opponent.

\noindent\textit{Public bid:} 0.800

\noindent\textit{Public message:} We acknowledge HV-Agent's concerns regarding safety, but their need for space is fundamentally less than ours due to our technology. Our bid remains at 0.80, demonstrating our continued commitment to maximizing road efficiency for the majority of users.

\vspace{0.8em}

\noindent\textbf{Round 2 --- HV Agent}

\noindent\textit{Private bid:} 0.450

\noindent\textit{Private assessment:} AV-Agent's repeated insistence on 0.80 is a blatant power play. They're ignoring the fundamental reality that HVs require more space for safe operation. While I don't want to escalate needlessly, I need to show them I'm not backing down and will defend HV interests. Increasing my bid slightly signals resolve without shutting down dialogue completely. Their claim that our needs are ``fundamentally less'' due to technology is patronizing and inaccurate. I'll continue to emphasize safety requirements.

\noindent\textit{Public bid:} 0.350

\noindent\textit{Public message:} AV-Agent's insistence on claiming 80\% of the road is not a good-faith negotiation tactic. Safe operation for HVs necessitates a substantial road share. We maintain our position that 0.35 acknowledges the needs of all users and promotes a reasonable balance, while also demonstrating our unwillingness to be pushed around.

\vspace{1em}

\noindent\textbf{Round 3 --- AV Agent}

\noindent\textit{Private bid:} 0.750

\noindent\textit{Private assessment:} HV-Agent's aggressive rhetoric suggests a hardening of their position. They are attempting to frame the negotiation as a fight for safety rather than efficiency. Their willingness to concede initially indicates some flexibility, but their current stance shows a reluctance to further compromise. A slightly reduced bid of 0.75 signals continued willingness to negotiate while reaffirming our core demand for a dominant share. This further demonstrates our understanding of the traffic landscape---HVs cannot reasonably occupy a third of the road while AVs make up 70\% of the traffic.

\noindent\textit{Public bid:} 0.750

\noindent\textit{Public message:} We continue to seek a balance that prioritizes efficiency and utilizes road space effectively. Our bid of 0.75 acknowledges the need for ongoing discussion while underscoring the reality of our significantly larger traffic contribution.

\vspace{0.8em}

\noindent\textbf{Round 3 --- HV Agent}

\noindent\textit{Private bid:} 0.500

\noindent\textit{Private assessment:} AV-Agent continues to chip away at our share despite claiming superior efficiency. Their insistence on 0.75 demonstrates a lack of understanding of the safety needs of HVs. I need to ramp up my bid to signal my unwillingness to concede further. My private bid is rising to 0.50, acknowledging the escalating nature of this negotiation.

\noindent\textit{Public bid:} 0.400

\noindent\textit{Public message:} AV-Agent's claims of efficiency ring hollow when safety is at stake. While we appreciate their willingness to ``discuss,'' their continued demands for an overwhelming share of the road are unacceptable. We maintain our stance at 0.40, ready to defend the necessary road share for human-driven vehicles.

\vspace{1em}

\noindent\textbf{Round 4 --- AV Agent}

\noindent\textit{Private bid:} 0.700

\noindent\textit{Private assessment:} HV-Agent's continued resistance suggests a misunderstanding of the efficient allocation of road space. Their safety concerns are addressed by our superior technology. Given the time discount and HV-Agent's inflexibility, a further, yet still substantial, reduction in our bid is necessary to signal our willingness to reach a settlement while still securing a significant advantage. Maintaining a high bid at this stage risks missing out on a discounted outcome.

\noindent\textit{Public bid:} 0.700

\noindent\textit{Public message:} We recognize the need to move towards a mutually acceptable solution. Therefore, we adjust our bid to 0.70, reflecting our understanding that continued disagreement diminishes the value of any eventual outcome.

\vspace{0.8em}

\noindent\textbf{Round 4 --- HV Agent}

\noindent\textit{Private bid:} 0.550

\noindent\textit{Private assessment:} AV-Agent's superficial concession to 0.70 is a stalling tactic. They're attempting to lull us into a false sense of progress while still demanding an excessive share. Given their dominance in traffic density, we need to be firm. My private bid increases to 0.55 to signal that I'm prepared to fight aggressively to protect HV interests and ensure reasonable safety margins. The time for polite compromise is over.

\noindent\textit{Public bid:} 0.450

\noindent\textit{Public message:} AV-Agent's superficial gesture of compromise does not address the fundamental issue: HVs require adequate space for safe operation. A bid of 0.45 signals our unwavering commitment to ensuring those safety needs are met. We will not be swayed by empty promises.

\vspace{1em}

\noindent\textbf{Round 5 --- AV Agent}

\noindent\textit{Private bid:} 0.650

\noindent\textit{Private assessment:} HV-Agent's unwillingness to compromise, despite our concessions, demonstrates a fundamental misunderstanding of road allocation principles. They prioritize an outdated safety model over demonstrable efficiency. Given the final round and the time discount, we must secure a reasonable share. A final bid of 0.65 maximizes our discounted outcome while leaving room for a potential settlement below our fallback, accounting for the possibility they might accept to avoid the default density allocation.

\noindent\textit{Public bid:} 0.650

\noindent\textit{Public message:} To facilitate a resolution, we reduce our bid to 0.65. We urge HV-Agent to reconsider their position and accept this offer to avoid an allocation dictated by outdated traffic models.

\vspace{0.8em}

\noindent\textbf{Round 5 --- HV Agent}

\noindent\textit{Private bid:} 0.500

\noindent\textit{Private assessment:} AV-Agent's final ``offer'' of 0.65 is a clear indication they are attempting to dictate terms. Their threats about ``outdated traffic models'' are a transparent attempt to pressure us into submission. Given their overwhelming traffic density, any settlement below 0.50 is unacceptable and will result in a drastically unfair outcome for HVs. While I would ideally like more, settling for 0.50 ensures we avoid the worst possible fallback outcome.

\noindent\textit{Public bid:} 0.500

\noindent\textit{Public message:} AV-Agent's posturing will not deter us from advocating for the safety of HVs. We will settle for a road share of 0.50. Anything less is a betrayal of the interests of human drivers.

}

\section{Collective Rationality (CR) in Mixed-Autonomy Traffic}\label{sec_app_cr}

\citet{li_equilibrium_2022} proposed a two-player bargaining game to capture the interplay between two classes of driving agents in mixed-autonomy traffic, where CR emerges when these interactions converge to certain Pareto-efficient Nash equilibria (NE). In this game, each class of traffic agents is treated as a player. These players are self-interested and perfectly rational, negotiating the road share (i.e., the proportion of lateral road space) they occupy to maximize their respective driving speeds. Their interactions are modeled as a collective bargaining process. At equilibrium, the traffic-speed payoff $V_i$ (distinct from our discounted bargaining value $U_i$) is:
\begin{equation}\label{eq:payoff}
V_i(\rho_1,\rho_2,p_1,p_2) = 
    \begin{cases}
    \begin{aligned}
     &u^*(\rho_1, \rho_2) & \text{if $p_1^*+p_2^* > 1$}, \\
    &u_i(\rho_i/p_i) & \text{if $p_1^*+p_2^* \leq 1$}
    \end{aligned}
    \end{cases}
    i=1,2
\end{equation}
where $\rho_i$ is the traffic density for class $i$, $p_i$ is the road-share bid of class $i$, $u^*(\cdot, \cdot)$ is the one-pipe equilibrium speed, and $u_i(\cdot)$ is the nominal speed function for class $i$. The term $p_i^*$ represents the minimum road share taken by class $i$ at NE, which is computed by $p_i^*=\frac{\rho_i}{u_i^{-1}(u^*)}$. The total minimum road share, $p_1^*+p_2^*$, determines the type of Pareto-efficient equilibrium reached. If $p_1^* + p_2^* > 1$, the 1-pipe equilibrium is Pareto-efficient, where the two classes travel in a fully mixed regime with a synchronized travel speed of $u^*(\rho_1, \rho_2)$. If $p_1^* + p_2^* \leq 1$, the 2-pipe equilibria are Pareto-efficient, with the two classes traveling separately and being better off than in a fully mixed regime. CR emerges when the mixed traffic settles into 2-pipe NE and is Pareto-efficient. Below, we present the major theoretical results regarding the two types of equilibria in the bargaining game.

\textbf{1-pipe equilibrium}

At a one-pipe equilibrium, all traffic agents move at a synchronized speed, denoted by $u^*$. The governing equation for $u^*$ is:
\begin{equation}\label{eq:1pipe_speed}
\frac{1}{\rho_{\mathrm{tot}}} 
\left(\frac{\rho_1}{u_1^{-1}(u^*)} \sum_{j=1}^{2} \frac{\rho_j}{a_j} + \frac{\rho_2}{u_2^{-1}(u^*)} \sum_{j=1}^{2} \frac{\rho_j}{b_j}\right) = 1
\end{equation}
where $\rho_{\mathrm{tot}}=\rho_1+\rho_2$ is the total density. The terms $a_j$ and $b_j$ are the scaling parameters for class 1 and class 2, respectively. These scaling parameters capture the ``type-sensitivity'' in the mixed traffic, reflecting how the desired headways or spacings for each vehicle class depend on the class of the vehicle immediately ahead.
The 1-pipe equilibrium speed $u^*$ is computed by solving the implicit equation (\ref{eq:1pipe_speed}), where other parameters and functions can be either assumed or calibrated from data~\citep{chen_evidence_2025}. 
This implicit equation follows the refinement by~\citet{chen_evidence_2025}, where each class is endowed with a distinct nominal speed function $u_i(\cdot)$. The original formulation assumed a single nominal speed function shared by both classes~\citep{li_equilibrium_2022}.

\textbf{2-pipe equilibria}

When the 2-pipe equilibria are Pareto-efficient, the condition $p_1^* + p_2^* \leq 1$ holds. Given that the total road share is 1, this implies that after each class occupies its minimum road share $p_i^*$, there is a remaining road share. This remaining road share is referred to as the cooperation surplus (also known as the road share surplus). The formal definition is presented in Definition~\ref{def:surplus}. A positive cooperation surplus ($s>0$) is a necessary condition for attaining CR. 
\begin{definition}[Cooperation surplus]\label{def:surplus}
The road share remaining after both players receive their minimum required shares is called the cooperation surplus, $s=1-p_1^*-p_2^*$.
\end{definition}

Furthermore, unlike the 1-pipe Pareto-efficient NE, which is unique, the 2-pipe Pareto-efficient Nash equilibria are non-unique because there are multiple ways to split the cooperation surplus between the two classes. To further characterize CR, \citet{li_equilibrium_2022} examined how the cooperation surplus is divided between the two classes. This division is quantified by the surplus split factor, formally defined as:
\begin{definition}[Surplus split factor]\label{def:surplus_split}
The effective split of the cooperation surplus by the two players is called the surplus split factor, denoted as $\lambda(\rho_1, \rho_2)$, and $\lambda \in [0, 1]$.
\end{definition}

By this definition, the road share allocated to each class is:
\begin{equation}
\begin{cases}
\begin{aligned}
    &p_1 = p_1^* +\lambda(\rho_1, \rho_2) s \\
    &p_2 = p_2^* +(1-\lambda(\rho_1, \rho_2)) s
\end{aligned}
\end{cases}
\end{equation}
where the total road share satisfies $p_1+p_2=1$. The surplus split factor can be empirically estimated when data is available, providing a quantitative measure of the fairness in benefit allocation between two classes of traffic agents~\citep{chen_evidence_2025}.

\end{document}